\documentclass[aps,prd,superscriptaddress,10pt,showpacs,nofootinbib,preprintnumbers,twocolumn]{revtex4} 
\usepackage[utf8]{inputenc}
\usepackage{graphicx}
\usepackage{dcolumn}
\usepackage{bm}
\usepackage{latexsym}
\usepackage{amsfonts}
\usepackage{amssymb}
\usepackage{amsmath}
\usepackage{color}
\begin{document}
\newcommand{\Fdiffs}{$\mathrm{Diff}_{\mathcal{F}}$~}%
\newcommand\red[1]{{\color{red}#1}}
\preprint{Imperial/TP/2016/AEG/2} 
\title{The uninvited guest in mixed derivative Ho\v{r}ava Gravity}
\author{Andrew Coates}
\affiliation{School of Mathematical Sciences, University of Nottingham, University Park, Nottingham, NG7 2RD, UK}
\author{Mattia Colombo}
\affiliation{School of Mathematical Sciences, University of Nottingham, University Park, Nottingham, NG7 2RD, UK}
\author{A. Emir G\"umr\"uk\c{c}\"uo\u{g}lu}
\affiliation{Theoretical Physics Group, Blackett Laboratory,
Imperial College London, South Kensington Campus, London, SW7 2AZ,
UK}
\author{Thomas P. Sotiriou}
\affiliation{School of Mathematical Sciences, University of Nottingham, University Park, Nottingham, NG7 2RD, UK}
\affiliation{School of Physics and Astronomy, University of Nottingham, University Park, Nottingham, NG7 2RD, UK} 
\date{\today}
\begin{abstract}
We revisit the mixed derivative extension of Ho\v{r}ava gravity which was designed to address the naturalness problems of the standard theory in the presence of matter couplings. We consider the minimal theory with mixed derivative terms that contain two spatial and two temporal derivatives. Including all terms compatible with the (modified) scaling rules and the foliation preserving diffeomorphisms, we calculate the dispersion relations of propagating modes. We find that the theory contains four propagating degrees of freedom, as opposed to three in the standard Ho\v{r}ava gravity. The new degree of freedom is another scalar graviton and it is unstable at low energies. Our result brings tension to the Lorentz violation suppression mechanism that relies on separation of scales. 
\end{abstract}
\maketitle

\section{Introduction}
 
The predictions of General Relativity (GR) are in perfect agreement with the currently available observations and experiments \cite{Will:2005va}. On the other hand, we have theoretical indications that GR might not be a complete theory; it is not perturbatively renormalizable and is thus expected to break down at high energies.

Ho\v{r}ava gravity  \cite{Horava:2009uw} exhibits improved behavior  at high energies due to the presence of higher order derivative terms in the action. If one insists on Lorentz invariance, higher order derivatives are known to lead to a breakdown of unitarity \cite{Stelle:1976gc}. However, Ho\v{r}ava gravity  is constructed in a preferred foliation, thus breaking local Lorentz symmetry. This property allows the space and time coordinates to have different scalings at high energies
\begin{equation}
t \to [k]^{-z} t \,,\qquad
x^i \to [k]^{-1} x^i\,.
\label{eq:horavascaling}
\end{equation}
As a result,  in $D+1$ dimensions the theory contains terms with $2$ time derivatives and at least $2z$ spatial derivatives. The minimum amount of scaling anisotropy that leads to power-counting renormalizability is $z=D$. The theory itself is defined by the invariance under foliation preserving diffeomorphism (\Fdiffs) symmetry given by
\begin{equation}
t\to t'(t),\quad x^i\to x^{i\,\prime}(x^i,t)\,.
\label{eq:fdiff}
\end{equation}
Collecting all terms invariant under \Fdiffs transformations the general action for the minimal theory ($z=3$) in 3+1 dimensions is given by \cite{Blas:2009qj} 
\begin{equation}
S=\frac{M_p^2}{2}\int  dt\,d^3x\,N\,\sqrt{g}\left(K_{ij}K^{ij}-\lambda K^2\right)+S_V\,,
\label{eq:horavaaction}
\end{equation}
where the ``kinetic'' terms are composed of the extrinsic curvature
\begin{equation}
K_{ij}=\frac{1}{2N}\left(\dot{g}_{ij}-\nabla_i N_j-\nabla_j N_i\right)\,,
\end{equation}
and  the action including the ``potential'' terms is
\begin{equation}
S_V \equiv \frac{M_p^2}{2}\int  dt\,d^3x\,N\,\sqrt{g}\left(\mathcal{L}_1+\frac{1}{M_*^2}\mathcal{L}_{2}+\frac{1}{M_*^4}\mathcal{L}_{3}\right)\,,
\label{eq:potential}
\end{equation}
where \(\mathcal{L}_{n}\) contains all terms invariant under (\ref{eq:fdiff}) which contain \(2n\) derivatives of the ADM variables $(N,g_{ij})$ [$N_i$ does not actually contribute]. In the UV, $k\gg M_*$, the higher derivative terms are expected to take over, resulting in modified dispersion relations $\omega^2 \propto k^6$. This provides an additional momentum suppression in the graviton propagators, and the theory is power-counting renormalizable \cite{Horava:2009uw, Visser:2009fg}. In the opposite regime $k\ll M_*$, the dispersion relations become relativistic, and the reduced IR theory has been shown to have regions in parameter space entirely consistent with observations  \cite{Blas:2009qj,Blas:2010hb,Audren:2014hza,Yagi:2013qpa,Frusciante:2015maa}. See also Ref.~\cite{Sotiriou:2010wn} for an early brief review.

Despite these attractive features, an open problem is to screen the Lorentz violations. Although the direct bounds on Lorentz violations in the gravity sector are weak, the bounds on Lorentz violating operators in the matter sector are very stringent \cite{Coleman:1998ti, Kostelecky:2008ts}. Even if one is willing to assume that lower-order Lorentz violating operators in the matter sector are absent at tree level, loop corrections will generate them and  fine tunings at order $10^{-20}$ would be needed to match experiments \cite{Collins:2004bp,Iengo:2009ix}. Moreover, observations require even the higher order Lorentz-violating operators to be suppressed in the matter sector \cite{Liberati:2012jf}.  Hence, preventing Lorentz violations from leaking from the gravity sector to the matter sector is an important issue.

Several ways to address this concern have been proposed in the literature. A symmetry enjoyed by all sectors may forbid lower-dimension Lorentz violating operators in the matter sector. Supersymmetry is one such example \cite{GrootNibbelink:2004za}, although this would require a supersymmetric version of Ho\v{r}ava gravity which is still unknown \cite{Xue:2010ih,Redigolo:2011bv,Pujolas:2011sk}. 
Another approach is to go beyond the perturbative realm, by strong interactions that take over at an intermediate scale between the Lorentz violation scale and some IR scale and accelerate the flow to Lorentz invariance in the IR \cite{Bednik:2013nxa,Kharuk:2015wga,Afshordi:2015smm}.

In this paper, we will instead focus on another potential resolution that was proposed in Ref.~\cite{Pospelov:2010mp}, where the Lorentz violating gravity sector is coupled to the Standard Model via power suppressed operators. This way the induced Lorentz violations in the matter sector scale as $\left(M_{*}/M_{p}\right)^2$ and can therefore be made small by regulating the relative size of $M_{*}$. 
However, the rather generic mechanism of Ref.~\cite{Pospelov:2010mp} is not entirely successful when applied to Ho\v{r}ava gravity. The obstruction is that non-dynamical vector gravitons do not undergo any modification with respect to GR,
leading to quadratic divergences that need to be fine-tuned away.\footnote{A more ambitious application of this mechanism was discussed in Ref.~\cite{Pospelov:2013mpa}, where Ho\v{r}ava gravity is coupled to supersymmetric matter for which  SUSY breaking is mediated by the Lorentz violations in the gravity sector. In this scenario, both the SUSY breaking and the Lorentz violations in the matter sector are controlled by the ratio $(M_*/M_p)^2$, i.e.~the suppression mechanism of \cite{Pospelov:2010mp} works in both ways. However, this scenario also requires that the graviton loop integrals are regulated by the higher order dispersion relations and hence its application to Ho\v{r}ava gravity requires taming the vector sector divergences.}
A way to remove this obstruction is to modify the behavior of the vector gravitons at high momenta. In Appendix \ref{app:vectormod} we argue that this issue cannot be resolved by adding higher order spatial derivatives to the `potential' part of the action. The authors of \cite{Pospelov:2010mp} proposed the addition of a single term \(\nabla_iK_{jk}\nabla^iK^{jk}\), which modifies the vector graviton sector at linear order while leaving the tensor and scalar dispersion relations qualitatively unchanged.\footnote{More precisely, the dispersion relation of the scalar mode \emph{does} change in the UV, but its momentum dependence stays the same, i.e. $\omega^2\propto k^6$.} Notably, this is a dimension $2z+2$ operator, beyond the truncation at $2z$. Moreover, it is not the only $2z+2$ dimensional operator and a possible concern is that additional operators can be generated by radiative corrections.

In order to address this concern, 
in Ref.~\cite{Colombo:2014lta},  the contributions of all terms of the form \(\left(\nabla_iK_{jk}\right)^2\) were studied. In this extension, all dispersion relations in the UV now become of the type $\omega^2 \propto k^4$. Although for the standard Ho\v{r}ava gravity, this is not enough for power-counting renormalizability, Ref.~\cite{Colombo:2014lta} argued that in the presence of mixed derivative terms, the UV scaling relation (\ref{eq:horavascaling}) is modified and for the new power--counting, these dispersion relations provide sufficient momentum suppressions in the amplitudes. Starting from modified anisotropic scaling rules, the fundamental basis for generic mixed derivative extensions were introduced in Ref.~\cite{Colombo:2015yha}, using a scalar field theory as an example. This new class of  Lifshitz-like (extensions to the Lifshitz scalar) theories are power-counting renormalizable and unitary.

Equipped with a consistent theoretical construction, the goal of the present paper is to apply the insights of  Ref.~\cite{Colombo:2015yha} to gravity and construct the most general mixed derivative extension of Ho\v rava gravity that includes all terms compatible with both the modified scaling rules and the  \Fdiffs symmetry. The resulting theory actually contains terms other than the \(\left(\nabla_iK_{jk}\right)^2\) terms considered in Ref.~\cite{Colombo:2014lta}. Excluding these new terms would require unjustified fine-tuning. However, a perturbative analysis reveals that they have a dramatic impact,  as they alter the dynamics by generating a new degree of freedom.

The rest of the paper is organized as follows. In Sec.~\ref{sec:Mixedderivative}, we briefly review the minimal mixed--derivative extension of Ho\v{r}ava gravity and construct the most general action that contributes at quadratic order in perturbations around flat spacetime. Sec.~\ref{sec:perturbations} is devoted to the calculation of dispersion relations for this theory and stability analysis. In Sec.~\ref{sec:Projectable}, we revisit this analysis by adopting the projectability condition. We conclude with Sec.~\ref{sec:discussion} where we discuss our results.

\section{Mixed derivative Ho\v{r}ava gravity}
\label{sec:Mixedderivative}

We start this section by reviewing the renormalizability and unitarity conditions for a mixed derivative extension of Ho\v rava gravity, first obtained in Ref.~\cite{Colombo:2015yha}. However, instead of working directly with a gravity theory, we resort instead to the simplified case of the Lifshitz scalar. This has been used in the literature in order to investigate the renormalization properties of standard Ho\v{r}ava gravity \cite{Visser:2009fg,Visser:2009ys} and this treatment was later extended  to include mixed derivative terms in Ref.~\cite{Colombo:2015yha}.

We focus on $3+1$ dimensions. To avoid the Ostrogradski instability the number of time derivatives is restricted to two. Moreover, we will only consider mixed derivative terms with  two time and two spatial derivatives. Hence we consider the 
following Lagrangian density for the free theory
\begin{equation}\label{actionLscalar}
\mathcal{L}_{\rm free}=\alpha \dot{\phi}^2+\beta \dot{\phi}(-\triangle)\dot{\phi} -
\sum_{\ell=1}^z \gamma_\ell \,\phi(-\triangle)^\ell\phi\,,
\end{equation}
where $\triangle \equiv \partial_i\partial^i$. 
In the UV, the terms with the coefficients $\beta$ and $\gamma_z$ dominate.  Hence, the theory exhibits the anisotropic scaling %
\begin{equation}
t\to [k]^{1-z}\,,\qquad
x^i\to[k]^{-1}\,.
\label{eq:newscaling}
\end{equation}
In Ref.~\cite{Colombo:2015yha} any self interaction with up to $2z$ derivatives was shown to be renormalizable provided
$z \ge 2\,$.
The minimum value of $z$  that satisfies this inequality, $z=2$ corresponds to relativistic scaling, as is clear from Eq.~(\ref{eq:newscaling}). This would mean that time and space derivatives scale the same way and the term $\ddot{\phi}^2$ is also allowed in the free Lagrangian, compromising unitarity. Requiring that unitarity is preserved imposes the following condition \cite{Colombo:2015yha}
\begin{equation}
z>2\,.
\end{equation}
Therefore, for a Lifshitz scalar theory with two temporal and two spatial derivative terms, self interactions with up to $6$ spatial derivatives are power-counting renormalizable provided that the free theory contains at least $6$ spatial derivatives.

We now proceed to construct a gravitational action that satisfies the same requirements. The action is of the form
\begin{equation}
S=\frac{M_p^2}{2}\int dt\,d^3x\,N \,\sqrt{g}\left(K_{ij}K^{ij}-\lambda K^2\right)+S_V+S_{\times}\,,
\label{eq:completeaction}
\end{equation}
where the kinetic terms with two time derivatives are built out of the extrinsic curvature, $K_{ij}$,
while the terms $S_V$, defined in Eq.~(\ref{eq:potential}), 
contain all operators compatible with the $\mathrm{Diff}_{\mathcal{F}}$ symmetry that have \(2\), \(4\) and \(6\) spatial derivatives, respectively. The last term in Eq.~(\ref{eq:completeaction}) is
\begin{equation}
S_{\times} = \frac{M_p^2}{2\,M_*^2}\int dt\,d^3x\,N \,\sqrt{g}{\cal L}_{\times}\,,
\end{equation}
which contains all \Fdiffs invariant operators that involve two spatial and two time derivatives.
The number of independent operators compatible with \Fdiffs and the power-counting is of order \(10^2\). However, below we are going to focus on linear perturbations around Minkowski spacetime, Hence, we only need to consider the terms that will contribute to the quadratic action in perturbation theory around this background. In this case, \(R_{ij}\), \(K_{ij}\) and \(a_i\) are all at least of linear order in perturbations, and so no term which is cubic (or higher) in these will survive the quadratic truncation. Furthermore, since the derivatives (excluding total derivatives) always enter with at least two perturbation order quantities, any terms related by commutation of derivatives are redundant at this order around Minkowski. Finally some terms are related, at this order in perturbation theory, by integration by parts. For example \(ND_i(Ra^i)\) is equivalent to \(-NRD_ia^i\) up to a total derivative and \(Ra_ia^i\), which is cubic order.

Following these criteria, we have significantly fewer terms to include in the action. The terms already present in  standard Ho\v{r}ava gravity (\ref{eq:potential}) are
\begin{align}
\mathcal{L}_{1}=&2\alpha a_ia^i+\beta R\,,\nonumber\\
\mathcal{L}_{2}=&\alpha_1RD_ia^i+\alpha_2D_ia_jD^ia^j+\beta_1R_{ij}R^{ij}+\beta_2R^2\,,\nonumber\\
\mathcal{L}_{3}=&\alpha_3D_iD^iRD_ja^j+\alpha_4D^kD_ka_iD_jD^ja^i\nonumber\\
&+\beta_3D_iR_{jk}D^iR^{jk}+\beta_4D_iRD^iR\,.
\end{align}
The relevant mixed derivative terms are
\begin{align}
\mathcal{L}_{\times}=&D_iK_{jk}D_lK_{mn}M^{ijklmn}
\nonumber\\
&+2\left(\sigma_1\mathcal{A}_i\mathcal{A}^i+\sigma_2\mathcal{A}_iD^iK+\sigma_3\mathcal{A}_iD_jK^{ij}\right)\,,
\label{eq:mixedaction}
\end{align}
where \cite{Colombo:2014lta} 
\begin{eqnarray}
M^{ijklmn}&\equiv&\gamma_1 \,g^{ij}g^{lm}g^{kn}+\gamma_2\,g^{il}g^{jm}g^{kn}\nonumber\\
&&+\gamma_3\,g^{il}g^{jk}g^{mn}+\gamma_4\,g^{ij}g^{kl}g^{mn} 
\end{eqnarray}
and 
\begin{align}
\mathcal{A}_i &\equiv\frac{1}{2N}\left(\dot{a}_i-N^jD_ja_i-a_jD_iN^j\right)\,,
\label{eq:extraterms}
\end {align}
is the \Fdiffs covariant combination which contains the time derivative of the acceleration. There is also a \Fdiffs covariant combination which contains the time derivative of the 3--curvature, namely\footnote{In the 4--d covariant formulation, the invariance of these quantities is more transparent. The two quantities can be defined in this case as 
\begin{equation}
\mathcal{A}_\nu \equiv -\frac{h^\mu_\nu}{2}\pounds_{u}a_\mu\,,\qquad
r_{\alpha\beta} \equiv -\frac{h^\mu_\alpha h^\nu_\alpha}{2}\pounds_{u}R_{\mu\nu}\,,
\end{equation}
where the Lie derivatives are along the normal vector $u_\mu$, the projection onto the constant time hypersurfaces is done through $h^\mu_\nu \equiv \delta^\mu_\nu + u^\mu u_\nu$,  $a_\mu$ and $R_{\mu\nu}$ are 4d covariant generalizations of the acceleration and the 3d Ricci tensor. In the ADM formulation, by replacing $u_\mu = \delta_\mu^0N$, the above definitions reduce to the ones given in Eqs.~(\ref{eq:extraterms}) and \eqref{eq:extraterms2}.}
\begin{align}
r_{ij} &\equiv \frac{1}{2\,N}\,\left(\dot{R}_{ij}-N^kD_kR_{ij} - R_{ik}D_jN^k-R_{jk}D_iN^k\right)\,.
\label{eq:extraterms2}
\end {align}
Naively, the terms $K^{ij}r_{ij}$ and $Kr$ are \Fdiffs scalars with the right number of derivatives and should be included in $\mathcal{L}_{\times}$. But as we show in Appendix \ref{app:proof}, they are redundant at the level of the action quadratic in perturbations around flat space time. 

As already discussed in Refs.~\cite{Pospelov:2010mp,Colombo:2014lta,Colombo:2015yha}, the mixed derivative terms in the first line of Eq.~(\ref{eq:mixedaction}) can be thought of as UV deformations of the kinetic terms of the tensor and scalar modes. However, we will see that the three terms on the second line (those involving \(\mathcal{A}_i\)) are instead related to a new scalar degree of freedom. That is, this theory has two tensor and two scalar degrees of freedom, in contrast with the three degrees of freedom in Ho\v{r}ava gravity.

\section{Perturbations around Minkowski}
\label{sec:perturbations}
We now perform the perturbative analysis of the theory given in Eq.~\eqref{eq:completeaction}. Since we focus on perturbations around flat spacetime, we adopt the following decomposition:
\begin{align}
N=1+A&,\quad N^i=B^i+\partial^iB\,,\nonumber\\
g_{ij}=\delta_{ij}(1+2\psi) + (&\partial_i\partial_j-\frac{\delta_{ij}}{3}\Delta)E+\partial_{(i}E_{j)}+\gamma_{ij}\,,
\label{eq:decomp}
\end{align}
where \(\partial^i\gamma_{ij}=\gamma^i_i=0\), leaving us with two degrees of freedom in the tensor sector. In the vector sector we have \(\partial^iB_i=\partial^iE_i=0\), leaving us with four degrees of freedom. Finally we have four scalar degrees of freedom, \(A, B, \psi\) and \(E\). This exhausts the ten degrees of freedom that can reside in a lapse $N$, a shift $N^i$ and a symmetric 3-metric (or a foliated 4-metric).

From here on, we shall proceed by expanding all perturbations in terms of plane waves, through
\begin{equation}
Q(t,\vec{x}) = \frac{1}{(2\pi)^{3/2}}\int d^3k \,Q_{\vec{k}}(t)\,e^{i\,\vec{k}\cdot\vec{x}}\,,
\end{equation}
where $Q(t,x^i)$ stands for any perturbation while $Q_{\vec{k}}(t)$ is the corresponding mode function. This operation non-trivially fixes the boundary conditions (see Ref.~\cite{Colombo:2014lta} for a discussion).
In the following, we will suppress the label $\vec{k}$ in order to lighten the  notation.
\subsection{Tensor sector}
Since the tensor modes are only affected by the first term in (\ref{eq:mixedaction}), the dispersion relations are the same as in Ref.~\cite{Colombo:2014lta}. Namely, the action quadratic in tensor perturbations is
\begin{equation}
S_{\rm tensor}^{(2)} = \frac{M_p^2}{8}\int dt \,d^3k \,\left(1+\gamma_2 \kappa^2\right)
\left(\vert\dot{\gamma}_{ij}\vert^2-\omega_T^2\vert\gamma_{ij}\vert^2\right)\,,
\end{equation}
where we defined $\kappa \equiv k/M_*$. The dispersion relation for the tensor perturbations is given by:
\begin{equation}
\omega_T^2 = k^2\,\frac{\beta -\beta_1 \kappa^2-\beta_3 \kappa^4}{1+\gamma_2 \kappa^2}\,.
\end{equation}
The linear stability of the tensor perturbations can be attained by requiring a positive kinetic term and a real frequency.
In the UV, i.e. $\kappa\gg1$, the kinetic term is dominated by the $\kappa^2$ part, which imposes $\gamma_2>0$. The dispersion relation in this regime is
\begin{equation}
\omega^2_{T} = -\frac{\beta_3 k^2}{\gamma_2} \left[\kappa^2+ {\cal O}(\kappa^0)\right]\,,
\end{equation}
requiring $\beta_3/\gamma_2<0$.

In the IR, i.e.~for $\kappa \ll 1$, the kinetic term is manifestly positive, so the only constraint comes from requiring a real propagation speed;
\begin{equation}
\omega^2_T = \beta\,k^2 \left[1 +{\cal O}(\kappa^2)\right]\,.
\end{equation}
Collecting all the conditions from stability of tensor modes at various scales, we have
\begin{equation}
\gamma_2 >0\,,\qquad
\beta_3 <0\,,\qquad
\beta>0\,.
\label{eq:stability-tensor}
\end{equation}

\subsection{Vector sector}

The original motivation for the mixed derivative extension of Ho\v{r}ava gravity is to overcome the technical naturalness problem in the suppression mechanism of Ref.~\cite{Pospelov:2010mp}. Although the four vector perturbations $B_i$ and $E_i$ correspond to two gauge modes and two non-dynamical modes, the gauge invariant combination $B_i-\dot{E}_i/2$ will still be generated virtually in graviton loops (like the Coulomb field in electromagnetism). However, in standard Ho\v{r}ava gravity, the vector propagator remains the same as in GR. As the suppression mechanism relies on loop integrals that are regulated in the UV, the vector loops lead to quadratic divergences. The addition of mixed derivative terms provides the necessary contribution to the vector propagator.

Considering that the quantity $\mathcal{A}_i$ in Eq.~(\ref{eq:mixedaction}) contains only scalar perturbations, the vector sector is only affected by the first term in (\ref{eq:mixedaction}). The action quadratic in vector perturbations thus coincides with the results of Ref.~\cite{Colombo:2014lta}:
\begin{equation}
S_{\rm vector}^{(2)} = \frac{M_p^2}{4}\int dt\,d^3k\, k^2\left[1+\frac{\kappa^2}{2}(\gamma_1+2\gamma_2)\right]
\left\vert B^i - \frac{\dot{E}^i}{2}\right\vert^2\,.
\label{eq:vecact}
\end{equation}
By specifying appropriate boundary conditions \cite{Colombo:2014lta}, the equation of motion for the non-dynamical $B^i$ field can be solved as
\begin{equation}
B^i = \frac{1}{2}\,\dot{E}^i
\label{eq:vecB}
\end{equation}
and therefore, by replacing this solution back in the action, we find that the action itself vanishes up to boundary terms. Hence, there are no dynamical vector modes, but the propagator now decays as $1/k^4$ in the UV.
\subsection{Scalar sector}\label{sec:scalarsector}
We can now proceed to studying the scalar sector of the theory, which is where the interesting features lie. The quadratic action for this sector is
\begin{widetext}
\begin{align}
S^{(2)}_{\mathrm{scalar}}=&\frac{M_p^2}{2}\int\mathrm{d}t\mathrm{d}^3k \left\{\left[3\left(1-3\lambda\right)+\left(\gamma_1+3\gamma_2+9\gamma_3+3\gamma_4\right)\kappa^2\right]\left|\dot{\psi}+\frac{k^2}{6}\dot{E}\right|^2+k^2\left(2\alpha+\alpha_2\kappa^2+\alpha_4\kappa^4\right)\left|A\right|^2\right.\nonumber\\
&~+2k^2\left[\beta+\left(3\beta_1+8\beta_2\right)\kappa^2+\left(3\beta_3+8\beta_4\right)\kappa^4\right]\left|\psi+\frac{k^2}{6}E\right|^2
+k^4\left[1-\lambda+\left(\gamma_1+\gamma_2+\gamma_3+\gamma_4\right)\kappa^2\right]\left|B-\frac{\dot{E}}{2}\right|^2\nonumber\\
&~+2k^2\left(\beta-\alpha_1\kappa^2+\alpha_3\kappa^4\right)\left[A^*\left(\psi+\frac{k^2}{6}E\right)+\mathrm{c.c.}\right]\nonumber\\
&~+k^2\left[1-3\lambda+\left(\gamma_1+\gamma_2+3\gamma_3+2\gamma_4\right)\kappa^2\right]\left[\left(B-\frac{\dot{E}}{2}\right)^*\left(\dot{\psi}+\frac{k^2}{6}\dot{E}\right)+\mathrm{c.c.}\right]\nonumber\\
&~+\frac{\sigma_1\kappa^2}{2}|\dot{A}|^2+\frac{k^2\kappa^2\left(\sigma_2+\sigma_3\right)}{2}\left[\left(B-\frac{\dot{E}}{2}\right)^*\dot{A}+\mathrm{c.c.}\right]
\left.+\frac{\kappa^2\left(3\sigma_2+\sigma_3\right)}{2}\left[\dot{A}^*\left(\dot{\psi}+\frac{k^2}{6}\dot{E}\right)+\mathrm{c.c.}\right]\vphantom{\left|\dot{\psi}+\frac{k^2}{6}\dot{E}\right|^2}\right\}\,.
\end{align} 
\end{widetext}

This action is manifestly gauge invariant as, at linear order, the quantities
\begin{equation}\label{eq:gaugeinvariants}
\Psi\equiv\psi+\frac{k^2}{6}E\,,\quad \mathcal{B}\equiv B-\frac{1}{2}\dot{E}\,,\quad\mathrm{and}\quad k\,A
\end{equation}
are invariant (hence do not transform) under \Fdiffs. Note that the perturbation $A$ is a scalar under 3-d diffeomorphisms, but under time reparametrizations of the type $t\to t+f(t)$, it transforms as $A\to A+f'(t)$. Therefore, the quantity $\partial_i A$ is gauge invariant while $A$ is not. That is, the gauge invariant plane wave mode function is $k\,A$.

We are left with three scalar degrees of freedom, two of which are dynamical. We can now use the momentum constraint to replace \(\mathcal{B}\), obtaining
\begin{align}
\mathcal{B}&=-\frac{1}{k^2}\frac{1}{1-\lambda+\left(\gamma_1+\gamma_2+\gamma_3+\gamma_4\right)\kappa^2}\nonumber\\
&~~\times
\left[\left[1-3\lambda+\left(\gamma_1+\gamma_2+3\gamma_3+2\gamma_4\right)\kappa^2\right]\left(\dot{\psi}+\frac{k^2}{6}\dot{E}\right)
\right.\nonumber\\
&\left.\qquad+\frac{\sigma_2+\sigma_3}{2}\kappa^2\dot{A}\right]\,.
\end{align}
Unlike the case in Ref.~\cite{Colombo:2014lta}, we can see that this time the field \(A\) is dynamical; for this reason we cannot perform any further reductions. We then have a scalar action with two dynamical degrees of freedom, \(Y=\left(\Psi,A\right)\), which can be written as
\begin{equation}
S^{(2)}_{\mathrm{scalar}}=\frac{M_p^2}{2}\int \mathrm{d}t\mathrm{d}^3k\left(\dot{Y}^{\dagger}K\dot{Y}-Y^{\dagger}MY\right)\,,
\label{eq:formalaction}
\end{equation}
where the matrices \(K\) and \(M\) are symmetric \(2\times 2\) matrices. The kinetic matrix \(K\) has components
\begin{align}
K_{11}=&\, 6+\left(4\,\gamma_1+6\gamma_2\right)\kappa^2+
\nonumber\\
&\frac{4+\left[8\left(\gamma_1+\gamma_2\right)+4\gamma_4\right]\kappa^2+\left[2\left(\gamma_1+\gamma_2\right)+\gamma_4\right]^2\kappa^4}{\lambda-1-\left(\gamma_1+\gamma_2+\gamma_3+\gamma_4\right)\kappa^2}\,,\nonumber\\
K_{12}=& -\sigma_3\kappa^2-\frac{\sigma_2+\sigma_3}{2}\frac{2\kappa^2+\left[2\left(\gamma_1+\gamma_2\right)+\gamma_4\right]\kappa^4}{\lambda-1-\left(\gamma_1+\gamma_2+\gamma_3+\gamma_4\right)\kappa^2}\,,\nonumber\\
K_{22}=& \frac{\sigma_1\kappa^2}{2}+\frac{\left(\sigma_2+\sigma_3\right)^2\kappa^4}{4\left[\lambda-1-\left(\gamma_1+\gamma_2+\gamma_3+\gamma_4\right)\kappa^2\right]}\,,
\end{align}
while for the mass matrix \(M\) we have
\begin{align}
M_{11}=& -2k^2\left[\beta+\left(3\beta_1+8\beta_2\right)\kappa^2+\left(3\beta_3+8\beta_4\right)\kappa^4\right]\,,\nonumber\\
M_{12}=& -2k^2\left[\beta-\alpha_1\kappa^2+\alpha_3\kappa^4\right]\,,\nonumber\\
M_{22}=&-k^2\left[2\alpha+\alpha_2\kappa^2+\alpha_4\kappa^4\right]\,.
\end{align}
The non-diagonal kinetic matrix can be diagonalized by performing a rotation to a new field basis $Z$ through
\begin{equation}
Z \equiv R^{-1}\,Y\,,
\end{equation}
with the rotation
\begin{equation}
R = \left(
\begin{array}{ll}
 1 & -\frac{K_{12}}{K_{11}}\\
 0 &1
\end{array}
\right)\,.
\end{equation}
In the new field basis, the kinetic matrix is diagonal $R^T K\,R = {\rm diag}(\bar{K}_1\,,\, \bar{K}_2)$ with eigenvalues 
\begin{equation}\label{kineigens}
\bar{K}_1=K_{11},\quad\bar{K}_2=\frac{\mathrm{det}\,K}{K_{11}}\,.
\end{equation}
It should be noted that this procedure is not unique. For instance, one could choose \(K_{22}\) and \(\mathrm{det}K/K_{22}\) for the kinetic eigenvalues, or adopt a basis obtained through an orthogonal rotation. However, the latter produces very complicated eigenvalues, rendering the treatment much more inconvenient. Provided that the rotation has non-zero determinant (i.e. the transformation can be inverted), the stability conditions are compatible. 

The first eigenvalue in Eq.~\eqref{kineigens} is independent of $\sigma_1$, $\sigma_2$, $\sigma_3$, while the second one vanishes when these parameters are zero. Hence, we identify the former mode as the scalar graviton of standard Ho\v{r}ava theory. In the IR the eigenvalues \eqref{kineigens} reduce to
\begin{equation}
\bar{K}_1=\frac{2\left(3\lambda-1\right)}{\lambda-1}+\mathcal{O}\left(\kappa^2\right),\quad\bar{K}_2=\frac{\sigma_1 \kappa^2}{2} +\mathcal{O}\left(\kappa^4\right)\,,
\end{equation}
leading to the following conditions for avoiding a ghost instability
\begin{equation}
\frac{3\lambda-1}{\lambda-1}>0\,,\quad\sigma_1>0\,.
\label{eq:noghost-sca}
\end{equation}

Thanks to the large number of UV relevant operators, there is more freedom to avoid high energy ghosts. In the $\kappa \gg 1$ limit, the kinetic eigenvalues become
\begin{align}
\bar{K}_1\!=\!&\left[2\left(\gamma_2+2\gamma_3\right)-\frac{\left(2\gamma_3+\gamma_4\right)^2}{\gamma_1+\gamma_2+\gamma_3+\gamma_4}\right]\kappa^2+\mathcal{O}\left(\kappa^0\right)\,,\nonumber\\
\bar{K}_2\!=\!&\left[\!\frac{\sigma_1}{2} \!-\!\frac{\left(2\gamma_1 \!+\!3\gamma_2\right)\sigma_2^2\!+\!2\left(\gamma_2\!-\!\gamma_4\right)\sigma_2\sigma_3\!+\!\left(\gamma_2\!+\!2\gamma_3\right)\sigma_3^2}{4\gamma_1\left(\gamma_2\!+\!2\gamma_3\right)\!+\!4\gamma_2\left(\gamma_2\!+\!3\gamma_3\right)\!+\!4\gamma_2\gamma_4\!-\!2\gamma_4^2}\!\right]
\nonumber\\
&\times\kappa^2+\mathcal{O}\left(\kappa^0\right)\,.
\label{eq:kineticIR}
\end{align} 

We finally obtain the dispersion relations. The equation of motion for the mode functions $Y$ can be obtained by varying the reduced action (\ref{eq:formalaction}) with respect to $Y^\dagger$
\begin{equation}
K\,\ddot{Y} + M \,Y = 0\,.
\end{equation}
We can then easily find the eigenfrequencies by considering a mode with $Y = Y_0 {\rm e}^{-i\,\omega\,t}$ and solving 
the equation
\begin{equation}
\mathrm{det}\left[\left(-i\omega\right)^2K+M\right]=0\,,
\label{eq:dispeq}
\end{equation}
which gives two distinct solutions for \(\omega^2\). The exact form of the dispersion relations are not suitable for presentation. For the present discussion, the expressions in the IR limit are instructive:
\begin{align}
\frac{\omega_1^2}{M_*^2}=&\frac{\beta\left(\beta-\alpha\right)\left(\lambda-1\right)}{\alpha\left(3\lambda-1\right)}\kappa^2+\mathcal{O}\left(\kappa^4\right)\,,\nonumber\\
\frac{\omega_2^2}{M_*^2}=& - \frac{4\alpha}{\sigma_1}
\nonumber\\
&\!\!\!
+\left[\frac{2\left[\beta\sigma_1\!-\!\alpha\left(3\sigma_2\!+\!\sigma_3\right)\right]^2}{\alpha\sigma_1^2\left(3\lambda-1\right)}\!-\!\frac{\left(\beta\sigma_1\!+\!2\alpha\sigma_3\right)^2}{\alpha\sigma_1^2}\!-\!\frac{6\alpha_2}{\sigma_1}\right]\frac{\kappa^2}{3}
\nonumber\\
&+\mathcal{O}\left(\kappa^4\right)\,.
\label{eq:dispersion}
\end{align}
We remark that the first expression retains the form of the IR dispersion relation for the scalar graviton in standard Ho\v{r}ava gravity, which upon imposing the stability of tensor modes (\ref{eq:stability-tensor}) and positivity of the kinetic terms (\ref{eq:noghost-sca}), retains the familiar condition 
\begin{equation}
 \beta>\alpha>0\,,
 \label{eq:gradient-stability-horava-scalar}
\end{equation}
to have a real propagation speed. On the other hand, the second mode has a tachyonic instability at leading order, i.e. a negative squared-mass. 
The time scale for this tachyonic instability is
\begin{equation}
t_{\rm ins} = \frac{\sqrt{\sigma_1}}{\pi\,M_*\,\sqrt{\alpha}}\,.
\end{equation}

\subsection{The scalar sector in the IR limit}

One might be tempted to assume that the higher dimensional mixed derivative operators (\ref{eq:mixedaction}) are UV deformations, irrelevant from the perspective of the low energy effective theory. However, from Eq.~(\ref{eq:dispersion}) we see that at  leading order, the dispersion relation of the second mode in the IR depends on the coupling constant $\sigma_1$ from a mixed derivative term. This is because the term $\mathcal{A}_i^2$ actually generates a kinetic term for an otherwise non-propagating perturbation in standard Ho\v{r}ava gravity. In that regard, the mixed derivative term  $\mathcal{A}_i^2$ is an IR relevant term as it provides the low energy kinetic term for the, now dynamical, lapse perturbation $A$. However, due to the two additional spatial derivatives in this term, the would-be gradient term $a^ia_i$ now provides a mass to $A$. 

It is therefore instructive to consider the IR theory and present a cleaner and more concise re-derivation of the perturbative dynamics. This will clearly describe the source of the new degree of freedom and the reason why it is either a ghost or a tachyon. We drop all the UV relevant terms such that the resulting action preserves the number of degrees of freedom of the full theory, obtaining
\begin{align}
S_{\mathrm{IR}}=\frac{M_p^2}{2}\int N\mathrm{d}t\sqrt{g}\mathrm{d}^3x\Bigg[&
K_{ij}K^{ij}-\lambda K^2+2\alpha a_ia^i
\nonumber\\
&+\beta R+\frac{2}{M_*^2}\sigma_1\mathcal{A}_i\mathcal{A}^i\Bigg]\,.
\end{align}
As we are interested only in the scalar sector of the theory, we fix the gauge and decompose the dynamical fields as
\begin{equation}
N=1+A\,,\quad N^i=\partial^iB\,,\quad g_{ij}=\delta_{ij}\left(1+2\psi\right)\,.
\end{equation}
Expanding the action up to quadratic order in perturbations, we arrive at the action
\begin{equation}
S^{(2)}_{\mathrm{IR, Scalar}}=\frac{M_p^2}{2}\int\mathrm{d}t\mathrm{d}^3x {\cal L}_{\rm IR}\,,
\end{equation}
with 
\begin{align}
\mathcal{L}_{\rm IR}=&-3\left(3\lambda-1\right)\dot{\psi}^2+\frac{\sigma_1}{2M_*^2}\nabla_i\dot{A}\nabla^i\dot{A}+2\left(3\lambda-1\right)\Delta B\,\dot{\psi}
\nonumber\\
&+2\beta\nabla_i\psi\nabla^i\psi+2\alpha\nabla_iA\nabla^iA+4\beta\nabla_iA\nabla^i\psi
\nonumber\\
&-\left(\lambda-1\right)\left(\Delta B\right)^2\,.
\end{align} 
Integrating out the non-dynamical mode $B$, the reduced action becomes
\begin{align}
\mathcal{L}_{\rm IR}=&\frac{2\left(3\lambda-1\right)}{\lambda-1}\dot{\psi}^2+\frac{\sigma_1}{2M_*^2}\nabla_i\dot{A}\nabla^i\dot{A}+2\alpha\nabla_iA\nabla^iA
\nonumber\\
&+4\beta\nabla_iA\nabla^i\psi+2\beta\nabla_i\psi\nabla^i\psi\,.
\end{align}
Due to the lack of kinetic mixing between \(A\) and \(\psi\) we can immediately read off the no-ghost conditions,
\begin{equation}
\frac{3\lambda-1}{\lambda-1}>0\,,\quad\sigma_1>0\,,
\end{equation}
as before. Furthermore, as the canonically normalized field is $\nabla_iA$,  the leading order contribution to the dispersion relation of this field comes from the second and third terms in the above action, allowing us to read off the mass of the massive mode as:
\begin{equation}
m^2=-\frac{4M_*^2\alpha}{\sigma_1}\,.
\end{equation}
Therefore, this IR exercise demonstrates that at leading order the unstable mode corresponds to the gradient of the lapse, i.e. $\nabla_i A$ which acquires a negative squared-mass. The remaining degree is massless and can be easily shown to correspond to the Ho\v{r}ava scalar.

\subsection{Changing the nature of the instability}

We have found above that the new scalar degree of freedom has a tachyonic instability, provided that the remaining stability conditions (\ref{eq:stability-tensor}), (\ref{eq:noghost-sca}) and (\ref{eq:gradient-stability-horava-scalar}) are satisfied. On the other hand, by relaxing one of these conditions, it is possible to obtain a real mass for the new degree of freedom. There are three ways to accomplish this: {\it i.~}For $\alpha<0<\beta$ the first scalar mode has a gradient instability; {\it ii.~}for $\beta<\alpha<0$ the tensor mode becomes a ghost;  {\it iii.~}for $\sigma_1<0$ the second scalar mode is a ghost. 

The limits on the parameters of the Ho\v{r}ava scalar and the tensor modes are well established \cite{Blas:2010hb,Yagi:2013qpa,Audren:2014hza}, so we will preserve the stability conditions for the modes already present in the standard Ho\v{r}ava theory. This leaves us with the third option. In fact, if we allow the IR effective theory to have a ghost with a mass larger than the cutoff of the low-energy action (strong coupling scale \cite{Papazoglou:2009fj,Kimpton:2010xi}), $M_{sc}$, then the ghost will not be generated in the regime of validity of the effective field theory \cite{Blas:2010hb}. This is an approach frequently used in effective field theories. However, here we actually know the UV completion of the theory, so we can eventually verify if the UV terms do indeed exorcise the ghost. 

For the IR effective theory to stay weakly coupled at all relevant scales one needs $M_*<M_{sc}$. This choice ensures that the higher derivative terms in the action become relevant before the IR theory becomes strongly coupled \cite{Blas:2009ck}. Then, the conditions for having a heavy ghost \emph{and} for avoiding strong coupling can be combined into one
\begin{equation}
\frac{4\,\alpha}{|\sigma_1|} > \frac{M_{sc}^2}{M_*^2} > 1\,,
\end{equation}
where we took $\sigma_1<0$. For the present discussion, we will assume $|\sigma_1| \ll \alpha$, which is necessary but not sufficient for satisfying the above conditions, although the details of our argument will not change in the case of a larger hierarchy between $M_{sc}$ and $M_*$.

From our previous analysis it is clear that the ghost  degree of freedom is not an artifact of some truncation (as is the usual assumption in effective field theories that contain a very massive ghost) but it actually continues to exist and propagate in the UV theory.  Hence, the only way to have positive energy at high momenta is if the kinetic term for this scalar changes sign at some intermediate momentum.
On the other hand, in the deep IR, the equation of motion for the new degree is, up to boundary conditions,
\begin{equation}
-\frac{|\sigma_1|}{2\,M_*^2} \ddot{A} - \alpha\,A =0\,.
\end{equation}
The coefficient of the kinetic term and the mass term have the same sign for  positive $\alpha$ and before  a canonical normalization. This suggests that when the former changes sign the latter should as well, else the scalar mode will turn from being a ghost to being classically unstable. 

\begin{figure}
\includegraphics[width=9.7cm]{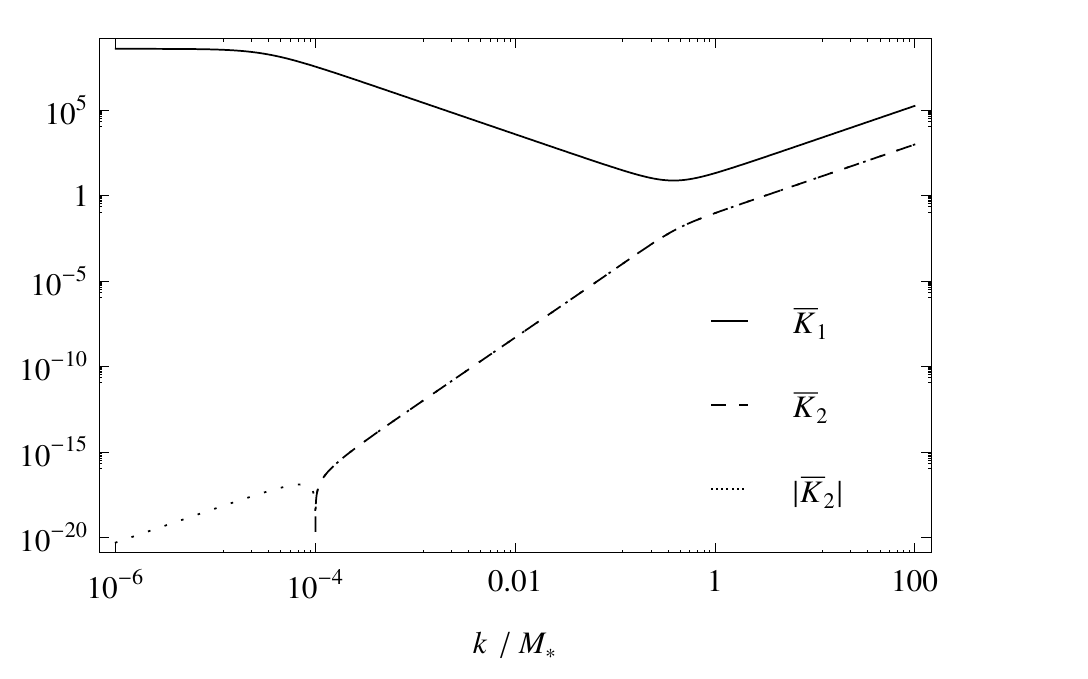}
\caption{The kinetic matrix eigenvalues (\ref{kineigens}) for the parameter set (\ref{eq:parameters}). The first eigenvalue (solid line) corresponds to the scalar graviton of Ho\v{r}ava gravity, while the second eigenvalue (dashed line, with absolute value shown as dotted line) is of the new degree arising from the mixed derivative extension. With the chosen parameters, the latter mode stops being a ghost at momenta $k\simeq 10^{-4}M_*$.}
\label{fig:kinetic}
\end{figure}

\begin{figure}
\includegraphics[width=9.7cm]{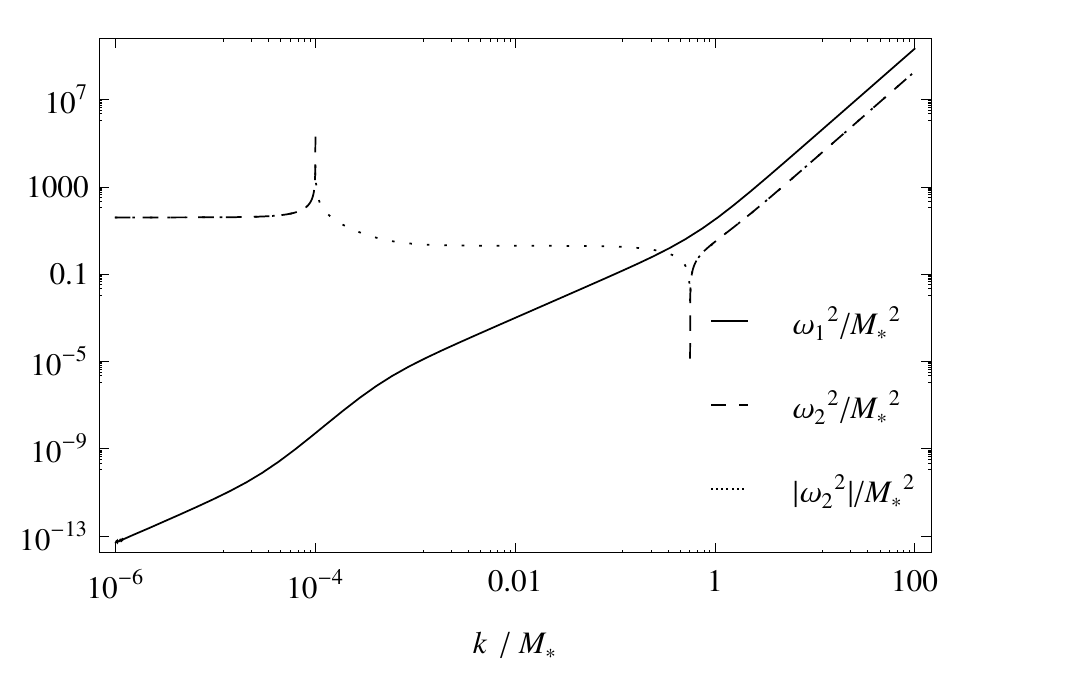}
\caption{The dispersion relation of the two modes, analytically obtained by solving Eq.~(\ref{eq:dispeq}) then evaluated using the parameter set (\ref{eq:parameters}). The solid line corresponds to the first (Ho\v{r}ava) mode, the dashed line corresponds to the new degree. The dotted line is the absolute value of the frequency of the second mode.}
\label{fig:dispersion}
\end{figure}

Clearly one needs to go beyond the IR limit of the dispersion relation in order to get the full picture.
To make this discussion concrete, we chose an example parameter set which is compatible with the current bounds on the IR parameters 
\begin{align}
& \alpha=10^{-7}\,,\;
\qquad
\beta-1=1.5\times10^{-7}\,,\;
\qquad
\lambda-1=10^{-8}\,,\nonumber\\
&\alpha_1=\alpha_2=\beta_1=\beta_2=-1\,,
\quad\;\;
\alpha_3=\alpha_4=\beta_3=\beta_4=-2\,,\nonumber\\
&
\gamma_1=\gamma_2=\gamma_3=1\,,
\; \gamma_4=-13\,,\;
\sigma_1 = -10^{-8}\,,\;
\sigma_2=\sigma_3=1\,,
\label{eq:parameters}
\end{align}
With these parameters, the standard Ho\v{r}ava scalar is stable both in the IR and UV, while the new mode is a heavy ghost in the IR and stable in the UV. In Fig.~\ref{fig:kinetic}, we show the kinetic terms for each mode as a function of momenta. The second mode is the new degree of freedom. Notice that at around $k\simeq10^{-4}M_*$, the sign of the kinetic term flips, and the mode becomes healthy again. This is due to the second term in $\bar{K}_2$ in Eq.~(\ref{eq:kineticIR}) becoming dominant.
In Fig.~\ref{fig:dispersion}, we show the dispersion relation as a function of the momentum. The first mode, i.e. the scalar graviton of Ho\v{r}ava theory has a dispersion relation $\propto k^2$ in the IR and $\propto k^4$ in the UV, as expected. The second mode starts off with a constant mass ($>M_*$), but when its kinetic term crosses zero and flips its sign the frequency of the mode diverges. It then experiences a tachyonic instability between momenta $10^{-4}M_* < k < M_*$. This implies that the theory is actually unstable at low-energies and the IR truncation that we used earlier to argue that the new scalar is a heavy ghost in the IR is simply misleading.


It seems likely that one could actually fine--tune the parameters of the theory so as to make the sign flip in the kinetic term exactly coincide with the one in the frequency and avoid  any instability at any momenta. The complexity of the full dispersion relations in the diagonal basis makes it particularly challenging to find such a tuning in practice. 
However, it is hard to imagine how it would be radiatively stable even if it exists.

\section{Invoking the Projectability Condition}
\label{sec:Projectable}
We now reexamine the results of the previous Sections by assuming further restrictions in the theory.
The issues associated with the unstable extra degree stem from the terms with coefficients $\sigma_n$, i.e those that contain time derivatives of the acceleration vector, which render the lapse dynamical. On the other hand, the projectability condition \cite{Horava:2009uw} constrains the lapse to be a function of time only. Hence, if one imposes this condition the offending terms will trivially vanish. In this restricted theory the lapse can be fixed by using the (space-independent) time reparametrization symmetry.\footnote{We remark that projectable Ho\v{r}ava gravity \cite{Horava:2009uw,Sotiriou:2009gy,Sotiriou:2009bx,Weinfurtner:2010hz} has recently been shown to  be renormalizable \cite{Barvinsky:2015kil}.} 

Imposing projectability affects only the scalar sector and the results in the previous section remain the same for the tensor and vector modes. Thus, the stability conditions for the tensor modes are still given by Eq.~\eqref{eq:stability-tensor} and the vector modes still acquire contributions from mixed derivative terms that improve the UV behavior.

The effect on the scalar sector is far more dramatic, as the projectability condition actually removes the second scalar mode. The coefficient of the kinetic term for the remaining scalar graviton is
\begin{align}
\bar{K}_{s,p}=&\, 6+\left(4\,\gamma_1+6\gamma_2\right)\kappa^2+
\nonumber\\
&\frac{4+\left[8\left(\gamma_1+\gamma_2\right)+4\gamma_4\right]\kappa^2+\left[2\left(\gamma_1+\gamma_2\right)+\gamma_4\right]^2\kappa^4}{\lambda-1-\left(\gamma_1+\gamma_2+\gamma_3+\gamma_4\right)\kappa^2}\,,
\label{eq:projkin}
\end{align}
while the dispersion relation is given by
\begin{equation}
\omega^2_{s,{p}}=\frac{-2\,\kappa^2\left[\beta+(3\,\beta_1+8\,\beta_2)\kappa^2+(3\,\beta_3+8\,\beta_4)\kappa^4\right]}{\bar{K}_{s,p}}\,.
\label{eq:dispersionproj}
\end{equation}
In the UV, the dispersion relation becomes $\omega^2_{s,p} \propto \kappa^4$, as expected from the modified scaling \eqref{eq:newscaling}. In the opposite limit, the IR expression for the coefficient of the kinetic term yields
\begin{equation}
\bar{K}^2_{s,p} = \frac{2\,(3\,\lambda-1)}{\lambda-1}\left[1+{\cal O}(\kappa^2)\right]\,,
\label{eq:projkinIR}
\end{equation}
while the dispersion relation reduces to
\begin{equation}
\omega^2_{s,p} = -\frac{\beta\,(\lambda-1)}{3\,\lambda-1}\,k^2\left[1+{\cal O}(\kappa^2)\right]\,.
\label{eq:IRproj}
\end{equation}
Requiring positivity of the kinetic term's coefficient \eqref{eq:projkinIR} in this limit yields:
\begin{equation}
\frac{3\,\lambda-1}{\lambda-1}> 0\,.
\end{equation}
Combining the above with the conditions from the tensor sector \eqref{eq:stability-tensor}, we see that the sound speed for the scalar mode is imaginary, leading to a gradient type instability.\footnote{In a cosmological setup, the amount of time necessary for the gradient instability to develop can be longer than the time scale of the Jeans instability, necessary for structure formation \cite{Mukohyama:2010xz}.} This is the well-known result of Ho\v{r}ava gravity with projectability condition \cite{Sotiriou:2009bx}.

In standard Ho\v{r}ava gravity, this IR gradient instability is accompanied by strong coupling in the limit $\lambda\to1$  \cite{Charmousis:2009tc,Blas:2009yd,Koyama:2009hc}.
This behavior emanates from the kinetic part of the action; the solution of the momentum constraint yields a shift vector with longitudinal component $B\propto (\lambda-1)^{-1}$. As the perturbative expansion of the action contains arbitrary powers of $B$, upon canonical normalization, terms of higher order acquire coefficients with increasing powers of the factor $(\lambda -1)^{-1}$.
Thus, if in the IR $(\lambda-1)$ runs to sufficiently small values from above, the perturbative expansion that led to the conclusion that there is an instability actually breaks down. This leaves open the possibility to have a non-perturbative restoration of the GR limit. Indeed, there are indications that $\lambda\to1$ limit is continuously connected to GR for spherically symmetric configurations \cite{Mukohyama:2010xz} and for cosmological solutions \cite{Izumi:2011eh,Gumrukcuoglu:2011ef}.\footnote{Around cosmological backgrounds, the reduced action for the dynamical degrees of freedom might even be compatible with perturbative expansion although there is no known local field redefinition to achieve this \cite{Gumrukcuoglu:2011ef}.} 

On the other hand, in the mixed derivative extension of projectable Ho\v{r}ava gravity, the scalar sector is modified. Although the gradient instability persists, the $\lambda\to1$ limit can still be perturbative. To be precise, the solution of the momentum constraint now gives (in the gauge $E=0$)
\begin{equation}
 B\Big\vert_{\lambda\to1} = \frac{1}{k^2}\,\frac{2-(\gamma_1+\gamma_2+3\,\gamma_3+2\,\gamma_4)\,\kappa^2}{(\gamma_1+\gamma_2+\gamma_3+\gamma_4)\kappa^2}\,\dot{\psi}\,,
\end{equation}
thus the longitudinal component of the shift vector no longer diverges in this limit. As a result, the strong coupling argument for projectable Ho\v{r}ava gravity does not apply to the mixed derivative extension and there is no indication that the perturbative expansion breaks down. However, the potential absence of strong coupling is not necessarily a blessing as the gradient instability at low momenta can no longer be screened.

A further implication of the finite $\lambda\to1$ limit arises in the dispersion relation for the Ho\v{r}ava scalar. In the original theory, the scalar dispersion relation is $\propto (\lambda-1)$ thus vanishes in this limit. On the other hand, the mixed derivative extension provides a finite contribution to the next to leading order term in \eqref{eq:dispersionproj}:
\begin{equation}
\omega^2_{s,p}\Big\vert_{\lambda \to 1} = k^2\left[
\frac{\beta\,(\gamma_1+\gamma_2+\gamma_3+\gamma_4)}{2}\,\kappa^2+{\cal O}(k^4)\right]\,,
\end{equation}
giving rise to a $k^4$ dispersion relation in the IR.

\section{Discussion}
\label{sec:discussion}

Coupling matter to gravity is an important challenge in Lorentz violating gravity theories. In particular, the main concern is to find a way to to avoid large Lorentz violating corrections to the matter sector, where Lorentz symmetry is extremely well constrained \cite{Kostelecky:2008ts}.

A mechanism which relies on separation of scales to suppress the Lorentz violating corrections was proposed in Ref.~\cite{Pospelov:2010mp}. However, adapting this mechanism to Ho\v{r}ava gravity introduces a technical naturalness problem in that the vector graviton loops diverge quadratically. It has been suggested in Ref.~\cite{Pospelov:2010mp} that adding one specific mixed derivative term could resolve this problem.  Mixed-derivative terms were studied in more generality in Refs.~\cite{Colombo:2014lta,Colombo:2015yha}. In Ref.~\cite{Colombo:2014lta} it was shown that theories with mixed derivative terms exhibit a modified scaling anisotropy and in Ref.~\cite{Colombo:2015yha} a tower of power-counting renormalizable, unitary Lifshitz--type theories were introduced.

In this paper, we applied the insights of Ref.~\cite{Colombo:2015yha} to gravity and introduced the
 minimal mixed-derivative extension of Ho\v{r}ava gravity, which includes all possible terms that are allowed by the new scaling and contribute to the quadratic action in perturbations around flat space.
The perturbative analysis of this more general version of the theory uncovered an instability, the nature of which depends on the choice of  parameters. In general, instead of the single scalar graviton appearing in Ho\v{r}ava gravity (and in the restricted mixed derivative theory of Refs.~\cite{Pospelov:2010mp,Colombo:2014lta}), there are actually two propagating scalar degrees of freedom. 
In the IR, the new scalar degree of freedom turns out to be either a tachyon or a ghost, {\em i.e.}~it has either imaginary mass or negative kinetic energy.

In the former case, the mode exhibits an exponential growth with a time scale
\begin{equation}
t_{\rm ins} = (7\,\times 10^{-31} {\rm s})\,\left(\frac{M_*}{10^9 {\rm GeV}}\right)^{-1}\left(\frac{\alpha}{10^{-7}}\right)^{-1/2}\,\sigma_1\,,
\end{equation}
where $M_*$ is the characteristic scale that suppresses higher order operators in Ho\v rava gravity, $\alpha$ is one of the parameters of the IR part of the actions, currently constrained to about 1 part in $10^{7}$ by weak field constraints \cite{Blas:2010hb}, and $\sigma_1$ is the coefficient of one of the terms that appear in the mixed derivative extension.
Attempting to render the instability inefficient  would require very large values of $\sigma_1$. 

 If instead the new scalar degree of freedom is a ghost, effective field theory wisdom suggests that its mass can be made to be heavy enough such that the instability is never reached within the regime of validity of the IR approximation. However, unlike most effective field theory treatments, we know that here the ghost is not a byproduct of the truncation and that this degree of freedom continues to propagate in the UV completion. Our analysis suggests that one cannot have a transition from a heavy ghost to healthy mode without fine--tuning. 

One way to avoid the unwanted scalar degree of freedom is to adopt the projectability condition of Ho\v{r}ava gravity. In this case the offending terms would be automatically excluded due to the restrictions in the field content (\(a_i=0\)).  However, in this case the known scalar degree of freedom is itself either a ghost or classically unstable, just as in the version without mixed derivative terms. Remarkably though, a preliminary analysis suggest that the mixed derivative terms  remove strong  coupling and make the projectable theory perturbative in the $\lambda\to 1$ limit.

Our results imply that adding mixed-derivative terms in order to address the naturalness problem found in Ref.~\cite{Pospelov:2010mp} has serious shortcomings. 
The mixed-derivative extension appears to be the only resolution without increasing the field content of the theory and so long as  \Fdiffs symmetry is preserved. An alternative would be to relax this symmetry in a way that allows the vector modes to be dynamical.  This is an interesting direction that will be explored in  future work.

\acknowledgments
The work of A.E.G. is supported by STFC grant ST/L00044X/1. The research leading to these results has received funding from the European Research Council under the European Union's Seventh Framework Programme (FP7/2007-2013) / ERC grant agreement n. 306425 ``Challenging General Relativity''. 

\appendix
\section{Modifying vector propagators in Ho\v{r}ava gravity}
\label{app:vectormod}

In this Appendix, we show that Ho\v{r}ava gravity with generic $z$ leads to the same linear equations for vector modes as GR. 
We start by considering linear vector perturbations around a Minkowski background
\begin{equation}
N=1\,,\qquad N^i = B^i\,,\qquad
g_{ij} = \delta_{ij} + \partial_{(i}E_{j)}\,,
\end{equation}
where scalar and tensor perturbations are ignored for the present discussion. Under infinitesimal transformation of spatial coordinates $x^i\to x^i+\xi^i$, we have
\begin{equation}
B_i-\frac{\dot{E}_i}{2}\to B_i-\frac{\dot{E}_i}{2}\,,
\label{eq:vectorgaugeinvariant}
\end{equation}
i.e. this combination involving transverse vectors is invariant. In fact, this is the only gauge invariant combination (up to a factor) one can construct out of vector fields. For this reason, any term in the action contributing only to one of $B_i$ or $E_i$ is expected to vanish at quadratic order. As an example, let us consider the spatial curvature tensor, which clearly does not depend on the shift vector (and hence its perturbation $B_i$),
\begin{align}
R_{ij} =& -\frac{\delta^{lm}}{2}\,\left[\delta g_{ij,lm}+\delta g_{lm,ij}-2\,\delta g_{l (i,j) m}
\right] \nonumber\\&+ \mathcal{O}({\rm perturbations}^2)\,.
\end{align}
Using the decomposition \eqref{eq:decomp}, it is immediate that the dependence on the transverse vector $E_i$ drops out at linear order in perturbations. Thus, we infer that any term in the action which contains two powers of the Ricci tensor will not contribute to the vector propagator. Similarly, the quantities $a_i$ and $R$ contain only scalar perturbations at linear order. Terms that mix these quantities with $R_{ij}$ will not contribute to the vector propagator due to 3d rotational symmetry of the Minkowski background.

Thus, any term in the action that can potentially modify the vector propagator should contain both $B_i$ and $E_i$ in the specific combination \eqref{eq:vectorgaugeinvariant}. The only such terms are the ones that involve Lie derivatives along the normal vector, e.g. the extrinsic curvature:
\begin{equation}
K_{ij} = -\left(\partial_{(i} B_{j)}-\frac{\partial_{(i}\dot{E}_{j)}}{2}\right)+ \mathcal{O}({\rm perturbations}^2)\,,
\end{equation}
where we only considered contributions to the vector sector. Notice that the trace of this quantity $K$, does not contribute to the quadratic vector action either.
Thus we have shown that in the action \eqref{eq:horavaaction}, only the $K_{ij}K^{ij}$ term contributes to the vector modes, independent of the number of spatial derivatives introduced by the $S_V$ term. 

If one insists on the symmetry \eqref{eq:fdiff} and the field content $N$, $N_i$ and $g_{ij}$, then there are only two ways to modify the quadratic action for vector modes with respect to GR: {\it i.~} to include higher powers of $K_{ij}$; {\it ii.~}to include terms quadratic in $K_{ij}$, but with spatial derivatives. Clearly, the former option involves more than two time derivatives and this is a threat for unitarity, so the only viable option is the latter one.

\section{Degenerate terms for linear perturbations}
\label{app:proof}
In this Appendix, we show that including the terms $K^{ij}r_{ij}$ and $Kr$ in $\mathcal{L}_{\times}$ would be redundant as, at quadratic order in perturbations around Minkowski space time, their contribution is no different than that of the $D_iK_{jk}D_l K_{mn}$ terms.

In the action (\ref{eq:mixedaction}), $r_{ij}$ is always combined with $K_{ij}$ whose leading order term is already linear in perturbations. Therefore, only the linear order term for $r_{ij}$ contributes to the action quadratic in perturbations.
From Eq.~(\ref{eq:extraterms}), we have
\begin{equation}
r_{ij} = \frac{1}{2\,N}\,\left[\dot{R}_{ij}-N^kD_kR_{ij} - R_{ik}D_jN^k-R_{jk}D_iN^k\right]\,.
\label{eq:rijdef}
\end{equation}
Since around flat spacetime, both $R_{ij}$ and $N_i$ are of order perturbations, only the first term in (\ref{eq:rijdef}) is of linear order. Explicitly,
\begin{equation}
 r_{ij} = \frac{1}{2}\left[\partial_k \dot{\Gamma}^k_{ij} - \partial_i \dot{\Gamma}^k_{jk}\right] + \mathcal{O}({\rm perturbations}^2)\,.
 \label{eq:rij-intermediate}
\end{equation}
From the definition of Christoffel symbols and the extrinsic curvature, we get 
\begin{align}
\dot{\Gamma}^k_{ij} =& D_j K_i^k+D_iK_j^k - D^kK_{ij} +D_iD_jN^k\nonumber\\
&+ \mathcal{O}({\rm perturbations}^2)\,,
\end{align}
using which, Eq.~(\ref{eq:rij-intermediate}) becomes
\begin{align}
r_{ij} =& \frac{1}{2}\left[D_i D^kK_{jk}+ D_jD^kK_{ik} - D_iD_jK -D_kD^k K_{ij}\right]\nonumber\\
&+ \mathcal{O}({\rm perturbations}^2)\,.
\end{align}
Notice that at leading order, the covariant derivatives commute and indices are raised/lowered by the flat Euclidean metric.

Finally, the combinations that appear in the action (\ref{eq:mixedaction}) can be written, up to boundary terms, as
\begin{align}
K^{ij} r_{ij}
\rightarrow& -D_iK^{ij}D^k K_{jk}
\nonumber\\&+\frac{1}{2}\left(D^k K^{ij}D_kK_{ij}+D_iK^{ij}D_jK\right)
\nonumber\\
&+ \mathcal{O}({\rm perturbations}^3)
\,,\nonumber\\
K\,r
\rightarrow& D_iK\,D^iK-D_iK^{ij}D_j K+ \mathcal{O}({\rm perturbations}^3)\,.
\end{align}

\end{document}